\documentclass[%
 aip,
rsi,%
 amsmath,amssymb,
reprint,%
longbibliography
]{revtex4-1}

\usepackage{graphicx}% Include figure files
\usepackage{dcolumn}% Align table columns on decimal point
\usepackage{bm}% bold math
\usepackage{xcolor}
\usepackage{amsfonts}
\usepackage{svg}

\newcommand\beq{\begin{equation}}
\newcommand\eeq{\end{equation}}
\newcommand\beqa{\begin{eqnarray}}
\newcommand\eeqa{\end{eqnarray}}

\begin{document}

%\preprint{AIP/123-QED}

\title{Chaos in Wavy-Stratified Fluid-Fluid Flow}% Force line breaks with \\

\author{Avinash Vaidheeswaran}
\affiliation{National Energy Technology Laboratory, Morgantown, WV, 26507, USA}
\affiliation{West Virginia University Research Corporation, Morgantown, WV 26506, USA} 

\author{Alejandro Clausse} 
\affiliation{CNEA-CONICET and Universidad Nacional del Centro, 7000 Tandil, Argentina}%

\author{William D. Fullmer}
\affiliation{National Energy Technology Laboratory, Morgantown, WV, 26507, USA}
\affiliation{AECOM, Morgantown, WV 26507, USA} %

\author{Raul Marino}
\affiliation{Universidad Nacional de Cuyo, 5500 Mendoza, Argentina}

\author{Martin Lopez de Bertodano}
\email{bertodan@purdue.edu}
\affiliation{School of Nuclear Engineering, Purdue University, West Lafayette, IN 47907, USA}%

\date{\today}% It is always \today, today,

\begin{abstract}
We perform a non-linear analysis of a fluid-fluid wavy-stratified flow using a simplified two-fluid model, i.e., the fixed-flux model (FFM) which is an adaptation of shallow water theory for the two-layer problem. Linear analysis using the perturbation method illustrates the short-wave physics leading to the Kelvin-Helmholtz instability (KHI). The interface dynamics are chaotic and analysis beyond the onset of instability is required to understand the non-linear evolution of waves. The two-equation FFM solver based on a higher-order spatio-temporal finite difference discretization scheme is used in the current simulations. The solution methodology is verified and the results are compared with the measurements from a laboratory-scale experiment. The Finite-Time Lyapunov Exponent (FTLE) based on simulations is comparable and slightly higher than the Autocorrelation function (ACF) decay rate, consistent with findings from previous studies. Furthermore, the FTLE is observed to be a strong function of the angle of inclination, while the root mean square (RMS) of the interface height exhibits a square-root dependence. It is demonstrated that this simple 1-D FFM captures the essential chaotic features of the interfacial behavior.
\end{abstract}

\maketitle

%\tableofcontents

%\begin{quotation}
%Multiphase flows--the simultaneous motion of material multiple constituent components which may be in different states of matter and are often characterized by an interface separating the components--are frequently observed in nature and industrial applications. Multiphase flows appear in a variety of geometrical configurations (flow regimes) depending on the properties of the constituents and the flow conditions \citep{Ishii2011}. Here, we study wavy-stratified flows of two immiscible liquids, i.e., separated by an interface, flowing counter-currently in a circular pipe. At flow rates beyond a critical value, the Kelvin-Helmholtz instability leads to a non-linear evolution of the fluid-fluid interface. We present our findings from the numerical analysis of a slightly inclined gasoline-water system, and compare the results with bench-scale experiment. We use a simplified two-fluid model--Eulerian-Eulerian model in which both phases are treated as inter-penetrating continuua--called the fixed-flux model and highlight the simple models' ability to predict the chaotic evolution of the interface. 
%\end{quotation}

\section{\label{sec:Intro}Introduction}
Applications involving fluid dynamics often exhibit nonlinear behavior. Chaos in single phase flows has been extensively studied in the past and still remains an area of active research. Transition to turbulence is a classic example where underlying strange attractors characterize the flow \cite{Lanford1982}. When an additional component or phase is present, new interfacial phenomena occur. Multiphase flows are known to exhibit hydrodynamic instabilities when differences in densities or velocities between the constituents exceed a critical limit. Chaos analysis of such systems is limited and includes the works of Pence and Beasley \cite{Pence1998} and Fullmer and Hrenya \cite{Fullmer2017}. In this article, we analyze the Kelvin-Helmholtz instability (KHI) which gives rise to well-known features including formation of clouds, waves in oceans and Saturn's magnetopause \cite{Wilson2012}. It is also important in industrial applications such as transport of oil and gas \cite{Alwahaibi2007}, and emergency core cooling systems of nuclear power reactors \cite{Duponcheel2016}.  
%Flow instabilities in two-phase flows arise due to differences in densities and/or velocities between the constituent phases. One such phenomenon is the Kelvin-Helmholtz instability (KHI) which is found in nature e.g., formation of clouds, waves in oceans and Saturn's magnetopause. It is also seen in industrial applications such as transport of oil and gas, and emergency core cooling system (ECCS) in nuclear reactors.

One of the first known measurements on the fluid-fluid interface dynamics was reported by Thorpe \cite{Thorpe1969} for a kerosene-water stratified flow in a rectangular channel. The experiments had a limited duration, sufficient to analyze the transition to wavy-stratified flow but not adequate for the subsequent evolution of waves. Recently, Duponcheel et al. \cite{Duponcheel2016} replicated the experiments of Thorpe \cite{Thorpe1969} and used the exact same geometry to obtain high-quality reliable data. In view of extending the analysis well past the inception of KHI, Vaidheeswaran et al. \cite{AV2016} performed experiments using gasoline and water in a round pipe. The post-processed data is presented here to further analyze the non-linear evolution of the waves. It is important to note that the conditions chosen for the experiment are such that the individual phases remain in the laminar flow regime. Hence, chaos arises from a multiphase interfacial instability and not from shear-induced turbulence. 
%This is essential to understand the non-linear stabilizing mechanism which ensures boundedness past the linear growth phase of KHI.\\
%The dominant wave modes obtained were in agreement with the predictions from stability analysis. 
%Numerical analysis pertaining to the study of interface dynamics is plentiful. 

Linear stability analysis (LSA) of the KHI is used to determine the mathematical nature of the system of equations. LSA can also help identify terms which stabilize the model and prevent immediate numerical divergence, i.e., terms that make the model well-posed. For instance, higher-order terms present in the TFM have been proven to influence the stability of the system of equations \cite{Ramshaw1978,Barnea1994,Taitel1976,MLB2013,Picchi2014,WDF2014} without completely hyperbolizing the model, i.e., no growth at all wavelengths. However, based on LSA of a well-posed TFM, there persists exponential growth at finite wavelengths, which is bounded by the inherent nonlinearity \cite{Krishnamurthy1992}. The nonlinear interaction between material shock waves and viscosity stabilizes the system by transferring growth at long wavelengths to dissipative short wavelengths.

In this work, we use a Fixed-Flux Model (FFM) \cite{MLB2017}, which is a simplified two-equation form of the TFM.  An advantage of the FFM is that the pressure gradient terms are not present and hence the numerical integration is straightforward. In Section \ref{sec:Expt} the water-gasoline experiments are described followed by the description of the FFM for stratified fluid-fluid flow in Section \ref{sec:TFM}. Stability of the hyperbolic-dispersive FFM is then discussed in the context of LSA and numerical simulation in Sections~\ref{sec:LSA} and \ref{sec:NLSA}. The work is concluded with a brief summary and outlook in Section~\ref{sec:Conclusions}. 

\section{\label{sec:Expt}Experiment}
\label{as}
In this work we compare numerical simulations to experiments conducted previously by \citet{AV2016}. The readers may also refer to \citet{MLB2017} for additional details of the experiment. Here, we briefly review the important features of the experiment for completeness. Additionally, a novel analysis of the experimental data is provided herein. The experiments were performed using an acrylic tube having diameter, $D=0.02 m$ and length, $L=2.4 m$. The test section is closed at both the ends and filled with water (heavier phase) and gasoline (lighter phase) having the following material properties: $\rho_{1}$=1000 $kg-m^{-3}$, $\rho_{2}$=720 $kg-m^{-3}$, $\mu_{1}$=0.001 $Pa-s$, $\mu_{2}$=0.0005 $Pa-s$, $\sigma$=0.04 $N-m^{-1}$. The subscripts 1 and 2 refer to water and gasoline respectively. The tube was held initially at rest to allow the mixture to settle. Once complete stratification occurs, the apparatus is tilted which initiates a counter-current flow pattern. 

\begin{figure}[tb] 
\begin{center}
\includegraphics[height=0.8\linewidth,trim=0.0cm 0.0cm 0.0cm 0.0cm, clip=true]{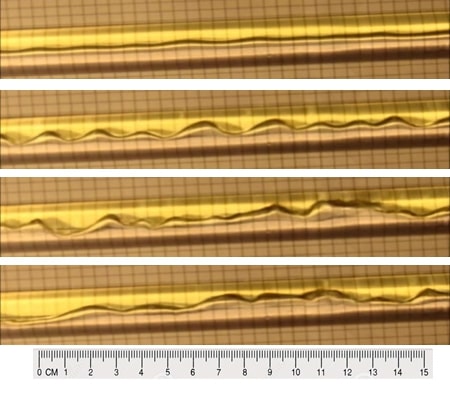}
\end{center}  
\caption{Evolution of wavy-stratified flow pattern of water and gasoline at 3.1$^{\circ}$ angle of tilt. Instantaneous snapshots correspond to (from top to bottom) $t$ = $0s$, $0.08s$, $0.32s$ and $8s$.}
\label{WaveEvolution}
\end{figure}

Figure \ref{WaveEvolution} shows instantaneous snapshots for a 3.1$^{\circ}$ angle of tilt. The initial wave-growth is qualitatively similar to the rectangular channel experiments of Thorpe \cite{Thorpe1969}. However, the following modifications to the design of Thorpe \cite{Thorpe1969} were made to better explore the chaotic behavior of the fluid--fluid interface: \emph{i}) increase the $L/H$ ratio to allow further development of waves and to obtain statistics for a longer duration, and \emph{ii}) reduce $D_{H}$ to ensure that the flow does not transition into turbulence. The chosen setup increases the duration of the experiments and isolates the chaotic behavior related to KHI from turbulence.

Preliminary findings from the experiments have been reported in \citet{AV2016} and \citet{MLB2017}. In this work, we extend the original experimental dataset by computing the Autocorrelation Function (ACF) of the wavy interface from the high speed video of the experiments. The images were digitized and the liquid level was detected using the Canny edge detector algorithm \cite{Canny1986}. The level signals at ten selected positions along the channel were post-processed to generate the best estimate for the ACF.

\section{\label{sec:TFM}Two-Fluid Model}

\begin{figure}[ht] 
\begin{center}
\includegraphics[height=0.5\linewidth,trim=0.0cm 0.0cm 0.0cm 0.0cm, clip=true]{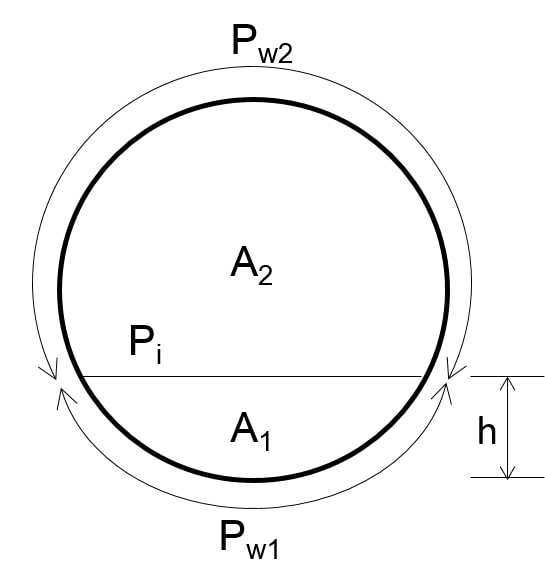}
\end{center}  
\caption{Geometrical representation of a pipe cross-section (diameter $D$) having heavier fluid at the bottom and lighter fluid at the top}
\label{Pipe_fig}
\end{figure}

The Eulerian TFM \cite{Ishii2011,Drew1999,Morel2015} provides a continuum framework to describe multiphase flow dynamics where the constituent phases are treated as inter-penetrating continua. Based on the dimensionality of the problem, further simplifications can be made. The 1-D TFM is derived from the 3-D Eulerian TFM by area-averaging over the channel cross-sectional area. The 1-D TFM may be further reduced to a FFM which is a variation of shallow water theory consisting of two coupled partial differential equations \cite{MLB2017}. As the term fixed-flux suggests, the simplification involves assuming a constant volumetric flux, $j = \alpha_{1} u_{1} + (1-\alpha_{1}) u_{2}$ where, $\alpha_{i}$ and $u_{i}$ represent phasic volume fraction and velocity. The 1-D FFM continuity and momentum equations for the wavy-stratified pipe flow depicted in Figure~\ref{Pipe_fig} follow \cite{MLB2017}:
\begin{equation}
\label{continuity}
\frac{\partial \alpha_1}{\partial t} + \frac{\partial}{\partial x}(\alpha_1 u_1) = 0 , 
\end{equation}
and 
\begin{equation}
\label{momentum}
\begin{split}
\frac{\partial u_1}{\partial t} + B_{21} \frac{\partial \alpha_1}{\partial x} + B_{22} \frac{\partial u_1}{\partial x} = C_{\alpha_{1}}(F_{\sigma} + F_{\nu} + F) , 
\end{split}
\end{equation}
respectively, where  
\begin{equation}
\label{constant}
C_{\alpha_{1}} = \frac{1-\alpha_1}{1-\alpha_1+r_{\rho} \alpha_{1}} .
\end{equation}
The convective coefficients are given by 
\begin{equation}
\begin{split}
\label{B21}
B_{21} = \Big[\Big(\frac{1}{1-\alpha_{1}} + \frac{1}{3}\Big)r_{\rho}(u_{2}-u_{1})^{2} - \frac{1}{3}(1+\rho)u_{1}^{2}\\ 
- \frac{\pi}{4}sin\Big[\frac{\pi}{2}\Big(\frac{1}{2} + \alpha_{1}\Big) \Big](1-r_{\rho})g_{y}D \Big] ,
\end{split}
\end{equation}
and
\begin{equation}
\begin{split}
\label{B22}
B_{22} = C_{\alpha_{1}}\Big[u_{1} + r_{\rho}\frac{\alpha_{1}}{1-\alpha_{1}}(2u_{2}-u_{1}) \\
+ \frac{2}{3}\frac{\alpha_{1}}{1-\alpha_{1}}(1-\alpha_{1}-r_{\rho} \alpha_{1})u_{1} \Big] .
\end{split}
\end{equation}
The forces on the right hand side (RHS) of Eq.~(\ref{momentum}) are due to surface tension, 
\begin{equation}
\label{ST}
F_{\sigma} = \frac{\sigma D}{2 \rho_1}\frac{\partial^{3}}{\partial x^{3}} \Big(1-cos\Big[\frac{\pi}{2}\Big(\frac{1}{2}+\alpha_{1}\Big)\Big]\Big) ,
\end{equation}
viscous diffusion, 
\begin{equation}
\label{VISC}
F_{\nu} = \nu_{T}\Big[\Big(\frac{1}{\alpha_1} + \frac{r_{\rho}}{\alpha_{2}} \frac{\partial}{\partial x} \Big(\alpha_{1} \frac{\partial u_{1}}{\partial x} \Big) + \frac{r_{\rho}}{\alpha_{2}}\Big(\frac{u_{1}}{\alpha_{2}}\frac{\partial \alpha_{1}}{\partial x} \Big)\Big] ,
\end{equation}
and wall and interfacial drag, 
\begin{equation}
\begin{split}
\label{F}
F = (1-r_{\rho})g_{x} - 16 \frac{P_{w1}\nu_{1}}{\alpha_{1}AD_{h1}} + 16 \frac{P_{w2}\nu_{2}}{\alpha_{2}AD_{h2}} \\ + \frac{1}{\alpha_1 \alpha_{2}}\frac{P_{i}\nu_{2}}{AD_{h2}}f_{i}r_{\rho}u_{2} .
\end{split}
\end{equation}
The interfacial friction factor, $f_i$, is adopted from the work of Andritsos and Hanratty \cite{Andritsos1987},
\begin{equation}
\label{fi}
f_{i} = 1 + 15 \sqrt{\alpha_{1}}\Big[\frac{u_{2}-u_{1}}{(u_{2}-u_{1})_{c}} - 1 \Big] .
\end{equation}

\subsection{\label{sec:LSA}Linear Stability Analysis}
\begin{figure}[!htb] 
\begin{center}
\includegraphics[height=0.8\linewidth,trim=0.0cm 0.0cm 0.0cm 0.0cm, clip=true]{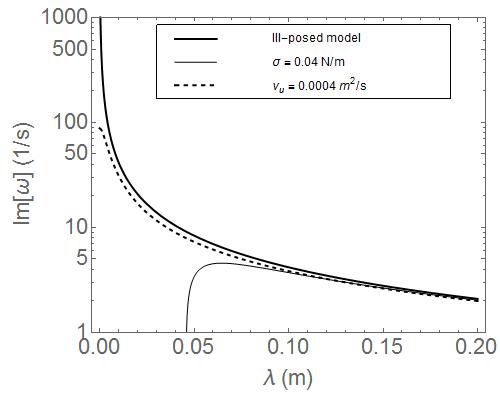}
\end{center}  
\caption{Dispersion relation for the FFMs using Equation \ref{DispEQ}. The curves correspond to the basic ill-posed model (unbounded exponential growth for $\lambda \to 0$), FFM with kinematic viscosity (bounded exponential growth for $\lambda \to 0$) and FFM with surface tension (well-posed).}
\label{Dispersion}
\end{figure}

Linear stability analysis of the FFM can be performed to determine the growth characteristics of different wave components. Accordingly, Equations \ref{continuity} \& \ref{momentum} are recast into the following vector form,
\begin{equation}
\label{systemEQ}
\textbf{A}\frac{\partial \underline{\phi}}{\partial t} + \textbf{B}\frac{\partial \underline{\phi}}{\partial x} + \textbf{D}\frac{\partial^{2} \underline{\phi}}{\partial x^{2}} + \textbf{E}\frac{\partial^{3} \underline{\phi}}{\partial x^{3}} = 0 ,
\end{equation}
where, $\underline{\phi}=[\alpha_{1} u_{1}]^{T}$. The force terms represented by Equation \ref{F} are neglected here because the focus is on analyzing the dynamic instability, rather than the kinematic instability \cite{MLB2017}. Perturbing the base state and linearizing Equation \ref{systemEQ} gives,
\begin{equation}
\label{DispEQ}
\mathbf{Det}[-i \omega \textbf{A} + i k \textbf{B} + (ik)^{2} \textbf{D} + (i k)^{3} \textbf{E}] = 0
\end{equation}
The solution to Equation~\ref{DispEQ} is the dispersion relation for growth rate $\omega$ as a function of wave number $k$.  Figure~\ref{Dispersion} shows the plots of magnitude of the imaginary component of $\omega$ as a function of the wavelength $\lambda$ = $2\pi/k$ for conditions beyond the critical KHI.
%using the following material properties: $D$=0.02 $m$, $\alpha_{1}$=0.54,$j$=0 $ms^{-1}$,$u_{1}$=0.092 $ms^{-1}$,$r_{\rho}$=0.72,$\nu_{1}$=$\nu_{2}$=40 $mm^{2}s^{-1}$ and $\sigma$=0.04 $N m^{-1}$. 

When the phasic viscosity or surface tension is neglected, the growth rate is unbounded as $\lambda \to$ 0. This constitutes the ill-posed condition beyond the KHI limit. This is the same result obtained using the 2-D Euler equations for a vortex sheet. Including kinematic viscosity removes the unboundedness (ill-posedness) \cite{Arai1980}, however, 
the zero wavelength growth rate remains the critical (most unstable) growth rate. In Figure~\ref{Dispersion}, $\omega_{i} (\lambda = 0) = 100$ rad/s so that appears quite similar to the ill-posed model over a majority of the frequency spectrum. Numerically, such a model would behave like the ill-posed model manifesting in an extremely grid-dependent instability (ill-behaved). On the other hand, surface tension renders the model well-posed and well-behaved with a finite valued cut-off wavelength, $\lambda_0$, below which the model is either neutrally stable (inviscid) or dissipative (viscous). The cut-off wavelength is approximately $\lambda_0 = 5$~cm for water-gasoline system. The results obtained are consistent with the work of Ramshaw and Trapp \cite{Ramshaw1978} who were the first to demonstrate the well-posed behavior of 1-D TFM using surface tension. Although the model including surface tension has been rendered well-posed, linear stability still indicates that finite wavelengths will grow at an exponential rate. In the following section, we turn to nonlinear analysis of numerical simulations of this linearly unstable model. The kinematic viscosity turns out to play a key role in the Lyapunov stability of the model.

\subsection{\label{sec:NLSA}Numerical Simulations}
Numerical simulations are performed using an in-house FFM solver originally developed for the prototypic Kreiss-Ystr\"{o}m Equations \cite{Fullmer2014}. The SMART scheme \cite{Gaskell1988} is used for spatial discretization and time marching is done by means of a strong stability preserving three-step, third-order Runge-Kutta method \cite{Gottlieb1998}. The numerical scheme reduces numerical diffusion (relative to first-order methods) while preserving numerical stability. 
%This would otherwise result in excess damping and eliminate the linear and non-linear dynamics pertaining to KHI.

\begin{figure}[tb] 
\begin{center}
\includegraphics[height=0.8\linewidth,trim=7.0cm 0.0cm 0.0cm 0.0cm, clip=true]{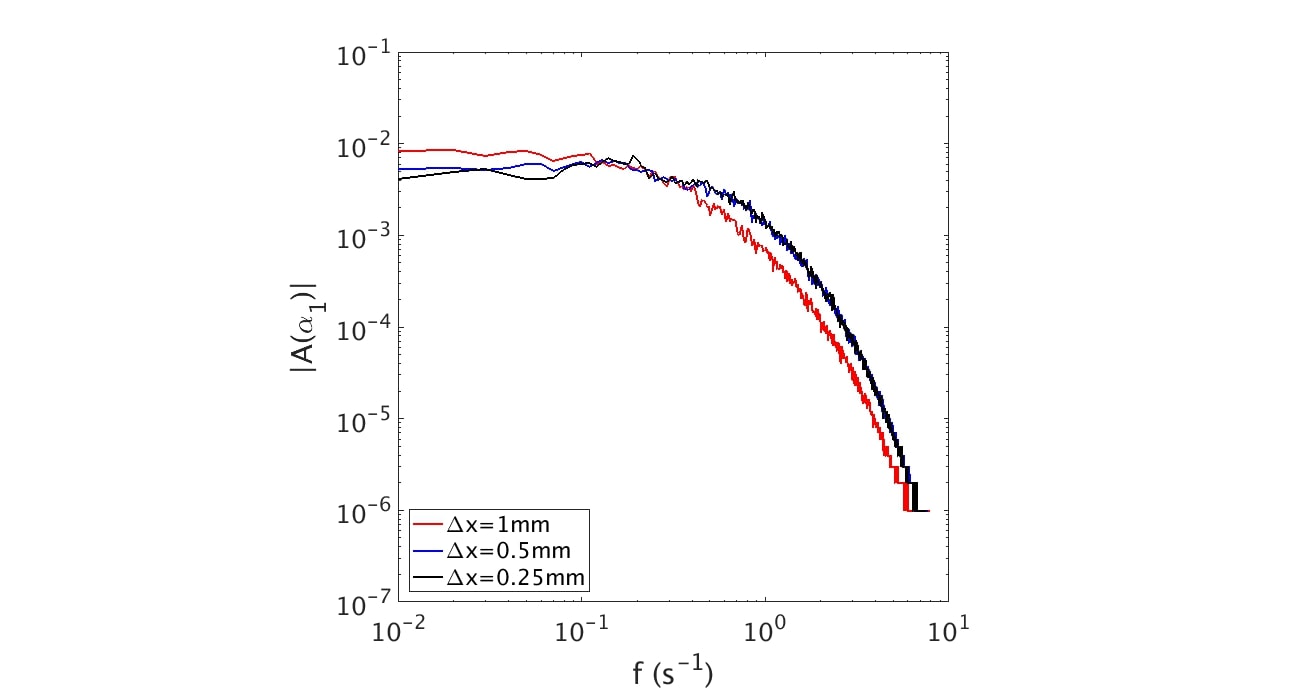}
\end{center}  
\caption{Fourier spectra evaluated from $\alpha_{1}$ time series based on successive grid refinements }
\label{Convergence}
\end{figure}

A spectral convergence test is used to ensure numerical stability, i.e., that the discretization scheme has not resulted in an ill-behaved numerical model. Three grid levels below $\lambda_0$ are used: $\Delta x$ = 1, 0.5 and 0.25~mm. The corresponding time step sizes are $\Delta t$ = 0.2, 0.1 and 0.05~ms, respectively. The domain length is the same as the pipe length used for the experiments but periodic boundary conditions are used to approximate the behavior in an infinite channel. This enabled us to perform numerical experiments over long periods of time in order to collect adequate statistics. The domain was initialized using a slightly perturbed liquid volume fraction field given by,
\begin{equation}
\label{IC}
\alpha_{1}(x,0) = \alpha_{1,0} + \delta\Big[\mathrm{exp}\Big(\frac{x-x_{01}}{\sigma_{01}}\Big)^{2} - \mathrm{exp}\Big(\frac{x-x_{02}}{\sigma_{02}}\Big)^{2} \Big]
\end{equation}
where, $\delta$ = 0.02, $\alpha_{1,0}$ = 0.542, $\sigma_{01} = \sigma_{02}$ = 0.02~m, $x_{01}$ = 0.45~m, and $x_{02}$ = 0.55~m. A constant velocity field is specified, $u_{1}$ = -0.1~m/s, calculated from the kinematic (quasi-static) condition, i.e., by solving $F = 0$ in Equation \ref{F}. The resulting FFT spectra in Figure~\ref{Convergence} shows that the long-wavelength (low frequency) energy generated from the linear KHI is transferred to higher frequencies through nonlinear dissipation, resulting in a continuous distribution of Fourier modes. The spectra corresponding to $\Delta x$ = 0.5mm and $\Delta x$ = 0.25mm indicate remarkable statistical convergence. Therefore, we use the more computationally affordable $\Delta x$ = 0.5mm for further analysis.

\begin{figure}[tb] 
\begin{center}
\includegraphics[height=0.7\linewidth,trim=0.0cm 0.0cm 0.0cm 0.0cm, clip=true]{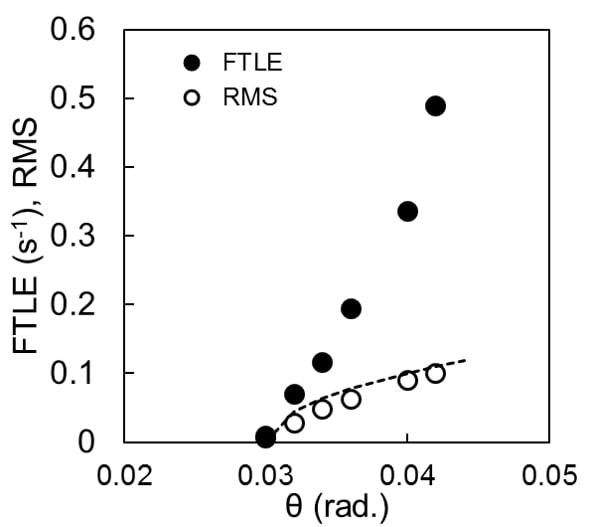}
\end{center}  
\caption{Finite Time Lyapunov Exponent and Root Mean Square evaluated using $\alpha_{1}$ time series. The dashed curve represents dependence proportional to $(\theta - \theta_{c})^{0.5}$.}
\label{FTLE}
\end{figure}

The Finite Time Lyapunov Exponent (FTLE) is approximated using nearby initial states \cite{Fullmer2014,MLB2017}. For a given angle of inclination, the initial condition, Equation \ref{IC} is perturbed by a small number, $10^{-8}$, approximately the square root of the double precision accuracy of the digital implementation. The L$_2$-norm of deviation in the trajectories relative to the (unperturbed) base state are calculated as a function of time and best-fit with an exponential curve to estimate the FTLE. This procedure is repeated at multiple angles of inclination. Figure~\ref{FTLE} shows that the FTLE has a strong dependence on $\theta$ beyond the onset of KHI. 
%It appears that chaos in the system considered and the degree of KHI are interdependent. 
With an increase in the angle of inclination, the relative velocity between the components increases making the fluid-fluid system more unstable. This occurs in conjunction with an increase in the magnitude of the FTLE. Also shown in Figure~\ref{FTLE} is the root mean square (RMS) of the liquid volume fraction which is a measure of the amplitude of sustained oscillations $\sim \theta - \theta_{c}$, where $\theta_{c}$ = 0.03 rad. is the critical angle when the system becomes chaotic. The second derivative of the RMS is negative which suggests that a supercritical type of bifurcation occurs at the $\theta_{c}$. Beyond approximately 2.4$^\circ$ or 0.042 rad., the waves become so chaotic that the interface reaches the top or bottom of the channel, leading to singularity in FFM.

\begin{figure}[tb] 
\begin{center}
\includegraphics[height=1.2\linewidth,trim=0.0cm 0.0cm 0.0cm 0.0cm, clip=true]{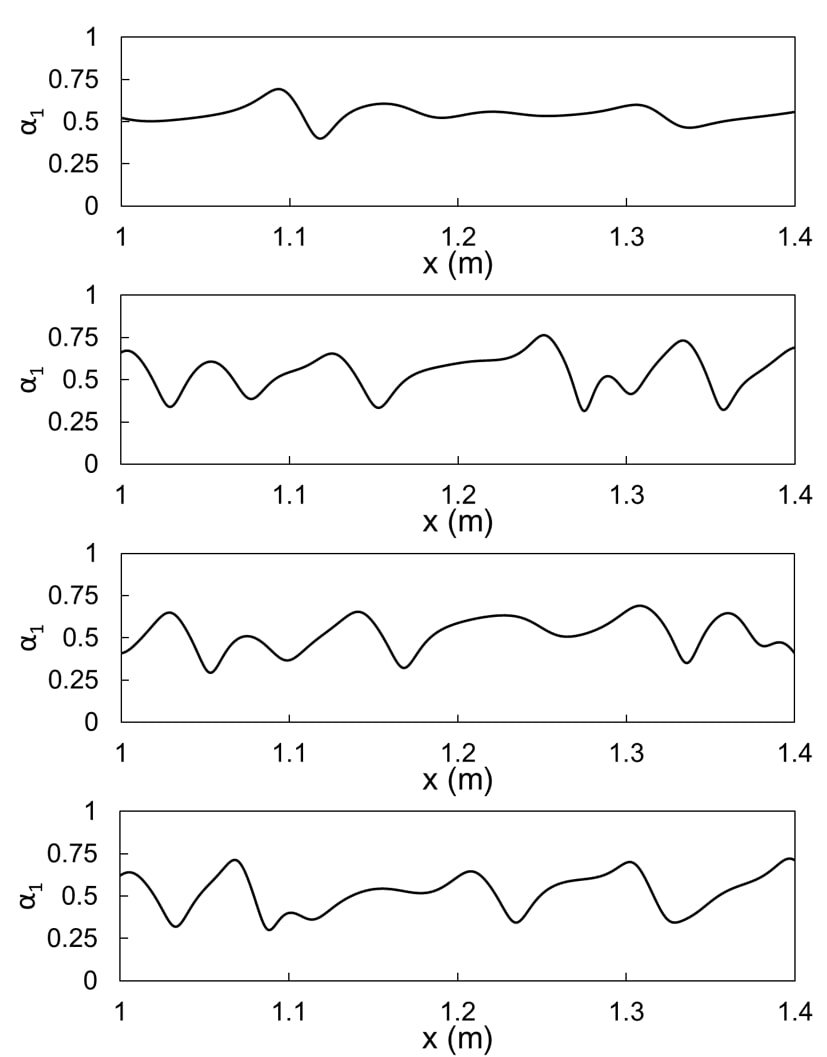}
\end{center}  
\caption{Instantaneous $\alpha_{1}$ distribution from FFM simulation at 2s, 4s, 8s and 15s (top to bottom)}
\label{wavelengths}
\end{figure}

Typical waveforms produced in the simulations are shown in Figure~\ref{wavelengths} for an inclination angle of 2.4$^\circ$.  Qualitatively, the waves are very similar to those observed experimentally, see Figure~\ref{WaveEvolution}. Quantitatively, however, the dominant wavelength of the simulation is approximately twice as large as observed in the experiments. The discrepancy is likely due to the fact that this is not a direct (``apples-to-apples'') comparison. In the experiment, the channel is suddenly tilted, the flow accelerates from rest and the channel ends fill up with single phase fluids. Because the FFM model is singular in the single-phase limit, the finite time experiment has been compared to an essentially infinite-time, periodic geometry at a lower inclination angle.

\begin{figure}[tb] 
\begin{center}
\includegraphics[height=0.75\linewidth,trim=0.0cm 0.0cm 0.0cm 0.0cm, clip=true]{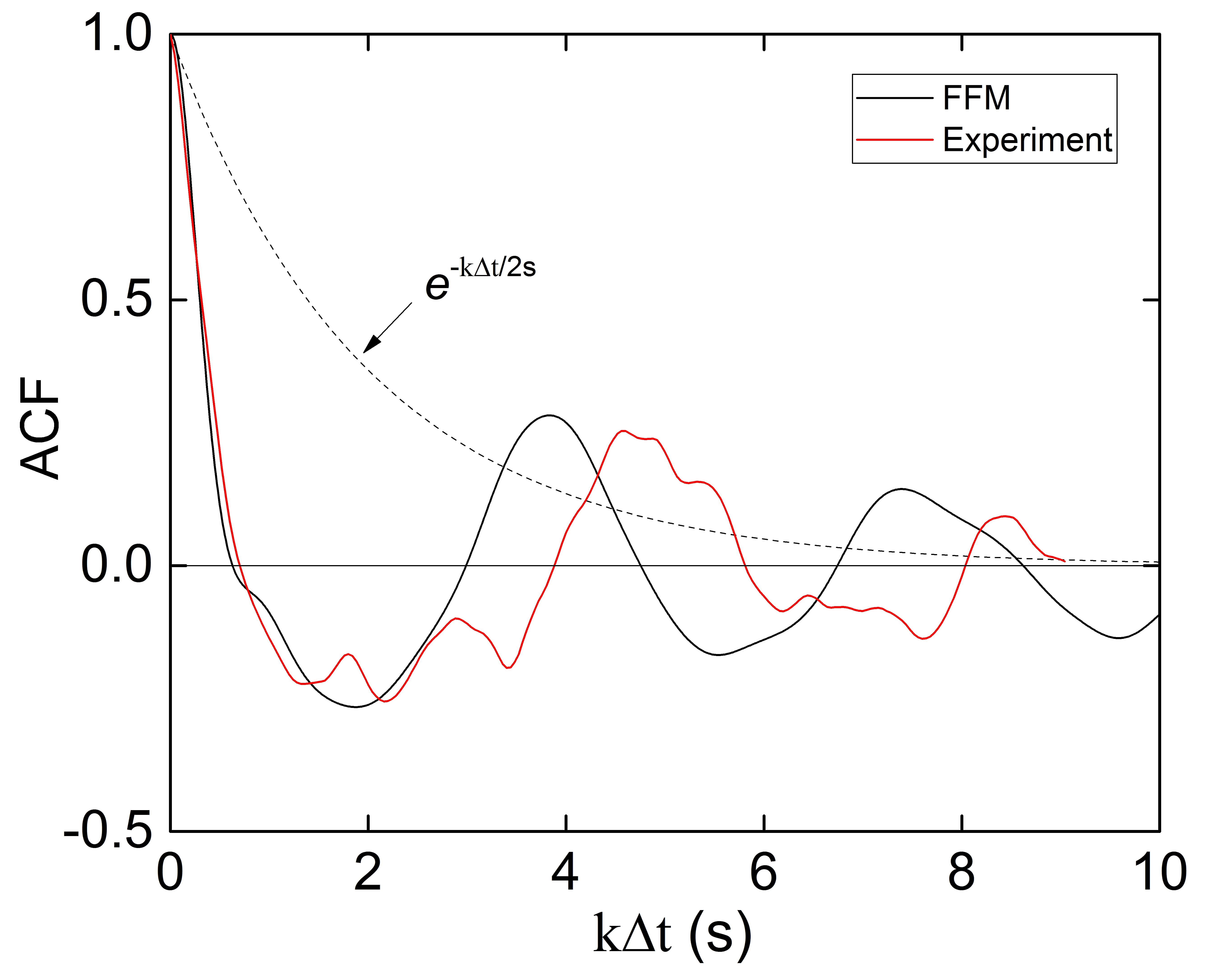}
\end{center}  
\caption{Comparison of ACF from experiments and numerical simulations. The dashed curve represents the exponential decay based on FTLE.}
\label{ACF}
\end{figure}

Beyond sensitivity to initial conditions (quantified by the positive values of FTLE), decay of correlations is another characteristic feature of chaos. Unlike FTLE, ACF is more easily obtained from physical experiments. Here, we compare the ACF using the discrete volume fraction signal from numerical simulations and experiments given by,

\begin{equation}
\label{eq:ACF}
\mathrm{ACF(k\Delta t)} = \frac{\textbf{E}\big[\alpha_{i},\alpha_{i+k}\big]}{\sqrt {\big[\sigma^{2}_{i} \sigma^{2}_{i+k} \big]}}
\end{equation}

\noindent where, $\sigma^{2}_{i}$ and $\sigma^{2}_{i+k}$ represent the variance of liquid volume fraction at two instants having a lag $k \Delta t$. Chaotic systems exhibit an exponential decay of the ACF characterized by a time scale related to the loss of predictability. While this rate of decay is not the same as the FTLE, it has been demonstrated that the FTLE and ACF are related for certain mathematical functions \cite{Badii1988,Julia2013}. More specifically, the rate of decay of correlation is bounded by the Largest Lyapunov Exponent in such cases. Figure \ref{ACF} shows the ACF measurements from the experiments and numerical simulations. The dashed curve in Figure \ref{ACF} represents this bound and it is seen that the relation \cite{Badii1988,Julia2013} holds true for this case of wavy-stratified flow as well.

\section{\label{sec:Conclusions}Conclusions}
The study highlights the ability of a 1-D FFM to capture the chaotic behavior of the interface in a wavy-stratified fluid-fluid flow. There appears to be a strong relationship between chaos in the fluid-fluid system and the degree of the KHI, so FFM simulations have been used to extend the scope of stability analysis past the linear theory. The numerical implementation of hyperbolic-dispersive numerical FFM is verified using a spectral convergence test, which is not commonly done for multiphase flows. The dynamics are reproduced, albeit at a slightly different angle of inclination of the experiment which may be attributed to factors including lack of modeling boundary layer development and its interaction with the interface, uncertainties in the constitutive models, and a periodic domain in the simulations compared to the finite domain of the experiments. However the chaotic evolution of the interface is captured by the model. The analysis revealed that infinitesimally perturbed states diverge based on the magnitude of the positive Lyapunov exponent before being bounded by a strange attractor. The FTLE is strongly dependent on the angle of inclination of the pipe while the RMS amplitude has a square-root dependence. Given the significant differences between numerical simulations and experiments, the predicted wave form shows qualitative resemblance even though the wavelengths are over-predicted. The decay rate of the ACF also seems to be comparable with the experiments.

%\section{References}
\bibliography{ref}

\end{document}